\documentstyle[12pt]{article}
\def\beq{\begin{equation}}   \def\eeq{\end{equation}}
\def\bea{\begin{eqnarray}}   \def\eea{\end{eqnarray}}
\def\be{\begin{equation}}   \def\ee{\end{equation}}

\newcommand{\matel}[3]{\langle #1|#2|#3\rangle}

\begin{document}

\begin{flushright}
UND-HEP-00-BIG\hspace*{.2em}11\\
\end{flushright}
\vspace{.3cm}
\begin{center} \Large
{\bf Heavy Flavour Physics: On Its More Than 50 Years Of
History, Its Future And The Rio Manifesto}
\footnote{Summary Talk given at HQ2K `Heavy Quarks at
Fixed Target 2000', Rio de Janeiro, Brazil, Oct. 9 -12,
2000}
\\
\end{center}
\vspace*{.3cm}
\begin{center} {\Large
I. I. Bigi }\\
\vspace{.4cm}
{\normalsize
{\it Physics Dept.,
Univ. of Notre Dame du
Lac, Notre Dame, IN 46556, U.S.A.} }
\\
\vspace{.3cm}
{\it e-mail address: bigi@undhep.hep.nd.edu }
\vspace*{0.4cm}

{\Large{\bf Abstract}}
\end{center}

After a reminder about how $\Delta S\neq 0$ physics has been
instrumental for the development of the Standard Model I sketch
theoretical technologies for dealing with nonperturbative QCD in
heavy flavour decays and state predictions for CP odd effects
as they were made in 1998. I review the exciting developments
in heavy flavour physics as presented at this conference. A central
message is presented in the `Rio Manifesto' where I recapitulate
the lessons we have learnt from charm physics, point out the
special role future dedicated charm studies can play in
revealing the presence of New Physics and give an
introduction to the relevant phenomenology focussed on
CP studies.


\tableofcontents

\vspace*{1.0cm}

\section{Introduction}

In the dawn of history, in a time long ago and a place far away,
in October 1946 Butler and Rochester saw a $V^0$ in their cloud chamber that
had been exposed to cosmic rays \cite{BUTLER}:
\beq
K^0 \to \pi ^+ \pi ^-
\eeq
This was the beginning of `heavy quarks at fixed target', and
things have never been the same again.
The comprehensive study of the dynamics of strangeness
has revealed the following fundamental features:
\begin{itemize}
\item
The $\tau -\theta $ puzzle led to the realization that parity is not
conserved in nature.
\item
The observation that the production rate exceeds the decay rate
by many orders of magnitude -- this was the origin of the
name `strange particles' -- was explained through postulating
a new quantum number -- `strangeness' -- conserved by the strong,
though not the weak forces. This was the beginning of the second
quark family.
\item
The absence of flavour-changing neutral currents was incorporated
through the introduction of the quantum number `charm', which
completed the second quark family.
\item
CP violation finally led to postulating yet another, the third
family.
\end{itemize}
All of these elements which now form essential pillars of the Standard
Model (SM) were New Physics at {\em that} time!
While strange quarks were thus instrumental for formulating
the Standard Model, charm quarks provided important
consistency checks; finally with the discovery of beauty quarks
(and $\tau$ leptons) the CKM ansatz became the standard
paradigm for CP violation with the only unknown
about the top quark being its mass.

The present status can be characterized as follows:
All six quark as well as lepton flavours have unambiguously
been discovered now.
A quite full SM profile is known about charm, the first predicted
flavour,and it happens to be a bit on
the dull side -- a point I will return to later. About beauty
quarks a lot is known, but its SM profile is not complete,
mainly because the latter is so exciting;
major efforts are under way to fill it out. Little is
directly known
about top quarks beyond their existence, their mass and their
affinity for beauty quarks; this will improve.

The SM forms a renormalizable and in that sense
self-consistent theory. It also works amazingly well.
Heavy quark and lepton flavours constitute essential elements of it --
and central mysteries:
\begin{itemize}
\item
Why is there a family
structure relating quarks and leptons?
\item
Why is there more than
one family, why three, is three a fundamental parameter?
\item
What is the origin of the observed pattern in the quark
(and lepton) masses
and the
CKM parameters? This pattern can hardly have come about by accident.
\item
Why are neutrinos massless -- or aren't they?
\end{itemize}
It certainly would represent an amazing triumph of the human
mind if pure thinking would lead us to the explanations
underlying these mysteries; however I am skeptical
about such an outcome. It seems much more likely that we
need to elicit more answers from nature through further
experimentation, however delphic nature's answers might turn out.
Dedicated studies of heavy flavour dynamics represent high
sensitivity searches for indirect manifestations of
New Physics.

The good news are that we are at a decisive period of flavour
physics:
\begin{itemize}
\item
{\em Direct} CP violation has been established in
$K_L$ decays.
\item
We are on the brink of observing the large CP asymmetries predicted
by the SM for certain $B$ decays.
\item
Studies of charm decays have finally reached a sensitivity
level where manifestations of New Physics could very realistically
show up.
\item
There is a good chance that at last leptons  are about to reveal
a nontrivial flavour structure.

\end{itemize}
My talk will be organized as follows: after
sketching the status of theoretical technologies for heavy flavour
physics in Sect. 2, I summarize the new results we have heard
about the dynamics of strange and beauty hadrons in Sects. 3
and 4, respectively; in Sect. 5 I present the `Rio Manifesto'
as a very central part of my talk, where I recapitulate the
glorious past of charm quarks and explain the reasons behind
my optimism that the best days of charm are still to come.

\section{Theoretical tools}

\subsection{QCD technologies of the 1990's}

Since we have to study the decays of quarks bound inside hadrons,
we have to deal with nonperturbative dynamics
\footnote{Since top quarks decay before they can hadronize,
their interactions can be treated perturbatively
\cite{DOK}.}
--
a problem that in general has not been brought under theoretical
control. Yet we can employ various theoretical technologies
based rather squarely on QCD
that allow to treat nonperturbative effects in special situations:
\begin{itemize}
\item
For {\em strange} hadrons where $m_s \leq \Lambda _{QCD}$
one invokes chiral perturbation theory.
\item
For {\em beauty} hadrons with $m_b \gg \Lambda _{QCD}$ one can
employ $1/m_b$ expansions in various incarnations; they should provide
us with rather reliable results, whenever an operator product expansion
can be applied \cite{HQT}.
\item
It is natural to extrapolate such expansions down to the charm
scale. This has to be done with considerable caution, though:
while the charm quark mass exceeds ordinary hadronic mass
scales, it does not do so by a large amount.
\item
Lattice QCD on the other hand is most readily set up at ordinary
hadronic scales; from those one extrapolates {\em down} towards the chiral
limit (which represents a nontrivial challenge) and
{\em up} to the charm scale and beyond.
\item
I want to direct your attention to the special position of charm. While
$1/m_c$ expansions extending from above
are a priori of somewhat dubious numerical
reliability, unquenched lattice calculations -- with some effort --
can reach the charm scale from below. For the later
discussion I want to note charm can thus take on a
central role as a bridge between different theoretical
technologies.

\end{itemize}
I do not consider quark models state-of-the-art
anymore in general. Yet as long as one remains aware of their
limitations and exercises good judgement, they can serve many important
purposes, one being to educate our intuition.  Yet for that to happen
one has to go beyond a naive formulation, as nicely demonstrated in
T\" ornkquist's talk \cite{TOERN}.

\noindent Let me add a few more specific comments:

Lattice QCD, which originally had been introduced to prove confinement
and bring hadronic spectroscopy under computational control is now making
major contributions to heavy flavour physics.
This can be illustrated with very recent results on decay constants
where the first {\em un}quenched results (with two dynamical
quark flavours) have become available \cite{KENWAY}:
\begin{itemize}
\item
\beq
f(D_s) =
\left\{
\begin{array}{l}
240 \pm 4 \pm 24 \; {\rm MeV} \; {\rm lattice\, QCD}\\
275 \pm 20 \; {\rm MeV} \;{\rm lattice\, QCD}\\
269 \pm 22 \; {\rm MeV}, \; {\rm world\, average\, of\, data
\, on \,}  D_s \to \mu \nu
\end{array}
\right.
\label{FDS}
\eeq
\item
\bea
f(B) &=& 190 \pm 6 \pm 20 ^{+9}_{-0} \; {\rm MeV},
\; {\rm lattice\, QCD}\\
f(B_s) &=& 218 \pm 5 \pm 26 ^{+9}_{-0} \; {\rm MeV},
\; {\rm lattice\, QCD}
\eea
\end{itemize}

The $1/m_Q$ expansions have become more refined and reliable
qualitatively as well as quantitatively:
\begin{itemize}
\item
The $b$ quark mass has been extracted
from data by three different groups following a pioneering study of
Voloshin \cite{VOLOSHIN};
their findings, when expressed in terms of the
socalled `kinetic' mass, read as follows:
\beq
m_b^{\rm kin} (1\, {\rm GeV}) =
\left\{
\begin{array}{l}
4.56 \pm 0.06  \; \;
{\rm GeV} \; \;  \cite{MEL}, \\
4.57 \pm 0.04  \; \;
{\rm GeV} \; \;  \cite{HOANG}, \\
4.59 \pm 0.06  \; \;
{\rm GeV} \; \;  \cite{SIGNER}
\end{array}
\right.
\eeq
The error estimates of 1 - 1.5 \% might be overly optimistic (as it
often happens), but not foolish. Since all
three analyses use basically the same input from the
$\Upsilon (4S)$ region, they could suffer from a common
systematic uncertainty, though.
\item
For the formfactor describing $B\to l \nu D^*$ at zero recoil
one has the following results:
\beq
F_{D^*}(0) =
\left\{
\begin{array}{l}
0.89 \pm 0.08  \; \;       \cite{URI1}, \\
0.913 \pm 0.042 \; \; \cite{BABARBOOK}, \\
0.935 \pm 0.03 \; \;  \cite{LAT} ,
\end{array}
\right.
\eeq
where the last number has been obtained in
lattice QCD.
\end{itemize}

There is a natural feedback between lattice QCD and $1/m_Q$
expansions: by now both represent mature technologies that
are defined in Euclidean rather than Minkowskian space;
they share some expansion parameters, while differing in others;
lattice QCD can evaluate hadronic matrix elements that serve
as input parameters to $1/m_Q$ expansions.

It has been accepted for a long time that heavy flavour decays
can serve as high {\em sensitivity} probes for New Physics. I feel
increasingly optimistic that our tools will be such that
that they will provide us even with high {\em accuracy} probes!

In describing nonleptonic two-body modes $B\to M_1 M_2$
valuable guidance has been provided over the years
by symmetry considerations
based on $SU(2)$ and to a lesser degree $SU(3)$.
Phenomenological models have played an important role
\cite{STECH,STECHNEUBERT}; more
often than not they involve factorization as a central
assumption. As already said, such models still play an important
role in
widening our horizon when used with common sense.
Yet the bar has been raised for them by the emergence of a new
theoretical framework for dealing with these decays. The
essential pre-condition for this framwork is the large energy
release, and it invokes concepts like `colour transparency'. While
those have been around for a while, only now they
are put into a
comprehensive framework. Two groups have presented results on
this \cite{BBNS,KLS}. The common feature in their approaches is that
the decay amplitude is described by a kernel containing the
`hard' interaction given by a perturbatively evaluated effective
Hamiltonion folded with form factors, decay constants and ligh-cone
distributions into which the long distance effects are lumped; this
{\em factorization} is symbolically denoted by
\beq
\matel{M_1M_2}{ H}{B} = f_{B\to M_1}f_{M_2}T^H * \Phi _{M_2} + ...
\eeq
The two groups differ in their dealings with the soft part:
KLS invoke Sudakov form factors to shield them against IR singularities.
In the BBNS approach on the other hand {\em no} IR singularities occur
to leading order in $1/m_b$. They do arise in nonleading orders where
they could be dealt with by introducing low energy parameters.
It is not surprising that the two groups arrive at different
conclusions: while BBNS infer final state interactions to be mostly
small in $B\to \pi \pi, K \pi$ with weak annihilation being suppressed,
KLS argue for weak annihilation to be important with final state
interactions {\em not} always small.

The trend of these results have certainly the ring of truth for me:
e.g., while factorization represents the leading effect in most cases
(including $B\to D\pi$), it is not of universal quality. One should
also note that the {\em non}-factorizable contributions move the
predictions for branching ratios towards the data -- a feature one could
not count on {\em a priori}. It is not clear to me yet whether the
two approaches are complementary or irreconcilable. Secondly one should view
these  predictions as preliminary: a clear disagreement with future data
should be taken as an opportunity for learning rather than
for discarding the
whole approach. This is connected with a third point: there are
corrections of order $\mu /m_b$ which are beyond our
computational powers. Since $\mu$ might be as large as 0.5 - 1 GeV,
they could be sizeable.

In summary: the theoretical technologies exist to describe two classes
of decays, namely fully inclusive transitions on one hand --
lifetimes, semileptonic widths, lepton spectra in
inclusive semileptonic
decays -- and the simplest exclusive modes -- semileptonic modes
with a single hadron or resonance in the final state and two-body
nonleptonic modes. Yet for all other modes
-- nonresonant three-body channels etc. -- all we have are models
of quite uncertain footing; it probably will take another breakthrough
to bring them under theoretical control.

\subsection{On Quark-hadron duality}

Most calculations are based on the concept of quark-hadron duality
(QHDu) in one form or another: when calculating
a rate on the quark-gluon level QHDu is invoked to equate the result
with what one should get for the corresponding process expressed in
hadronic quantities.

This concept is very successful for processes initiated by hard
dynamics: $e^+e^- \to \; hadrons$ well above thresholds, the widths for
$Z^0 \to b \bar b$ or $c \bar c$ jets or $W\to c \bar s$ jets etc.
For the purpose of treating beauty and charm decays two new more
specific questions arise in applying $1/m_Q$ expansions:
(i) How low can the scale be for QHDu
to still apply?
(ii) QHDu {\em cannot} be exact. The question is how large are actually
the uncertainties that enter here \cite{SHIFMAN,OSAKA}.

The $1/m_Q$ expansion should certainly work well for beauty, yet for charm
the situation is much more iffy, and realistically
one can hope for no more than semiquantitative results there.
QHDu represents an approximation the quality of
which is process-dependant and increases with the amount of
averaging or `smearing' over hadronic channels.

There is a lot of
folklore concerning QHDu, but no general theory.
That is not surprising: for QHDu can be addressed in a
quantitative fashion only {\em after} nonperturbative
effects have been brought under control, and that has happened only
recently in beauty decays.
Considerable insight exists into the physical origins of QHDu
violations: (i) They are caused by the exact location of
hadronic thresholds that are notoriously hard to evaluate.
Such effects are implemented through `oscillating terms'; i.e.,
the fact that innocuous, since suppressed contributions
${\rm exp}(-m_Q/\Lambda )$ in Euclidean space turn into
dangerous while unsuppressed sin$(m_Q/\Lambda )$ terms in
Minkowski space. (ii) There is bound to be some sensitivity
to `distant cuts' \cite{OPTICAL}. (iii) The validity of the
$1/m_c$ expansion arising in the description of
$B\to l \nu D^*$ is far from guaranteed.

Based on general considerations and analyses in model field theories
like the
't Hooft model (QCD in 1+1 dimensions with
$N_C \to \infty$) one can say the following: while one would not be
overly surprised if {\em nonleptonic} decays even of beauty hadrons
were to exhibit significant violations of QHDu, one has
good reason to be
confident that {\em inclusive semileptonic} decays of beauty hadrons are
described by $1/m_Q$ expansions quite accurately. This is further
re-inforced by the successful treatment of inclusive
$\tau$ decays where an accurate value for the strong coupling
$\alpha _S$ has been
extracted in full agreement with determinations at high energies like
at $Z^0$.

\subsection{CKM matrix -- present and future}

PDG2000 lists the following values for the CKM matrix
elements as 90\% C.L. ranges:
\beq
|V_{CKM}| =
\left(
\begin{array}{ccc}
0.9750 \pm 0.0008 & 0.223 \pm 0.004 & 0.003 \pm 0.002 \\
0.222 \pm 0.003 & 0.9742 \pm 0.0008 & 0.040 \pm 0.003 \\
0.009 \pm 0.005 & 0.039 \pm 0.004 & 0.9992 \pm 0.0002
\end {array}
\right)
\label{CKM3}
\eeq
{\em Without} imposing three-family unitarity
the numbers in particular
for the top couplings are much less restrictive:
\beq
|V_{CKM}| =
\left(
\begin{array}{cccc}
0.9735 \pm 0.0013 & 0.220 \pm 0.004 & 0.003 \pm 0.002 & ...\\
0.226 \pm 0.007 & 0.880 \pm 0.096 & 0.040 \pm 0.003 & ...\\
0.05 \pm 0.04 & 0.28 \pm 0.27 & 0.5 \pm 0.49 & ... \\
... & ... & ... & ...
\end {array}
\right)
\label{CKM4}
\eeq

PDG2000 quotes
\footnote{Since there are common uncertainties in $|V(td)|$ and
$|V(ts)|$ this ratio is not obtained directly from Eq.(\ref{CKM3}).}
\beq
\left| \frac{V(ts)}{V(td)} \right| > 4.17
\eeq
as inferred from the lower bound on $\Delta m(B_s)$.
Very
recently somewhat higher values have been stated:
$|V(ts)/V(td)| > 4.6$.

An optimist might say that there is intriguing though not
conclusive evidence that the first glimpse of $B_s$ oscillations
has been caught with $\Delta m(B_s) \sim 17.5 \; {\rm ps}^{-1}$. If so,
D0 and CDF should have little trouble establishing it soon
\cite{DUCCIO}.

Theoretically the cleanest way to extract $|V(td)|$ is measuring the
branching ratio for $K^+ \to \pi ^+ \nu \bar \nu$. Within the
SM one predicts \cite{BUCHBURAS}
\beq
0.5 \cdot 10^{-10} \leq
{\rm BR} (K^+ \to \pi ^+ \nu \bar \nu) \leq
1.2 \cdot 10^{-10}
\eeq
BNL experiment E789 has seen one celebrated event corresponding to
BR$(K^+ \to \pi ^+ \nu \bar \nu) = 1.5 ^{+3.4}_{-1.2}
\cdot 10^{-10}$. Its final sample will have a single event sensitivity
of about $0.7 \cdot 10^{-10}$; its successor, E 949, is expected
to yield around 10 events for the SM branching ratio \cite{JAIN}.
At Fermilab the experiment CKM has been proposed to measure
$K^+ \to \pi ^+ \nu \bar \nu$ at the 100 event level
\cite{NGUYEN}. These are truly heroic efforts!

I would like to add three comments here:
\begin{enumerate}
\item
The brandnew CLEO number for $|V(cb)|$ from
$B\to l \nu D^*$ --
$|V(cb)F_{D^*}(0)| = (42.4 \pm 1.8 \pm 1.9)\times 10^{-3}$ --
falls outside the 90\% C.L. range stated by PDG2000
for the expected values of
$F_{D^*}(0)$.

\item
The OPAL collaboration has presented
a new {\em direct} determination of $|V(cs)|$ from
$W\to H_c X$: $|V(cs)| = 0.969 \pm 0.058$
\cite{OPAL}.
\item
Using these values one finds
\beq
|V(ud)|^2 + |V(us)|^2 + |V(ub)|^2 = 1.000 \pm 0.003 \; ,
\eeq
which is perfectly consistent with the unitarity of the
CKM matrix. Yet using instead $|V(ud)| =
0.9740 \pm 0.0005$ as extracted from
nuclear $0^+ \to 0^+$ transitions, one obtains \cite{TOWNER}
\beq
|V(ud)|^2 + |V(us)|^2 + |V(ub)|^2 = 0.9968 \pm 0.0014 \; ,
\eeq
i.e., a bit more than a 2 $\sigma$ deficit in the unitarity
condition.

\end{enumerate}

With these input values for the CKM parameters
one can make predictions on CP asymmetries,
at least in principle and to some degree. I will confine myself
to a few more qualitative comments.
\begin{itemize}
\item
If there is a single CP violating phase
$\delta$ as is the case in the KM
ansatz one can conclude based on the $\Delta I = 1/2$ rule:
$\epsilon ^{\prime}/\epsilon \leq 1/20$. The large top mass
-- $m_t  \gg M_W$ -- enhances the SM prediction for $\epsilon$
considerably more than for $\epsilon ^{\prime}$ for a
given $\delta$ and therefore on quite general grounds
\beq
\epsilon ^{\prime}/\epsilon \ll 1/20 \; .
\eeq
\item
Of course the KM predictions made employed much more
sophisticated theoretical reasoning. Before 1999 they tended to
yield -- with few exceptions
\cite{FABB} -- values not exceeding $10^{-3}$
due to sizeable cancellations between different contributions.
\item
Once the CKM matrix exhibits the {\em qualitative}
pattern given in Eq.(\ref{CKM3}),
it neccessarily follows that certain $B_d$ decay channels will
exhibit CP asymmetries of order unity. To be more specific one can
combine what is known about $V(cb)$, $V(ub)$, $V(ts)$ and $V(td)$
from semileptonic $B$ decays, $B_d - \bar B_d$ oscillations and
bounds on $B_s - \bar B_s$ oscillations with or without using
$\epsilon$ to construct the CKM unitarity triangle
describing $B$ decays. A crucial question to which I will return
later centers on the proper treatment of theoretical uncertainties.
A typical example is \cite{PARODI}:
\beq
{\rm sin} 2\phi _1[\beta] = 0.716 \pm 0.070 \; , \;
{\rm sin} 2\phi _2[\alpha] = - 0.26 \pm 0.28
\eeq
\end{itemize}

\section{CP violation in strange and in beauty
transitions}

\subsection{$\Delta S \neq 0$}

Direct CP violation has now been established
in $K_L$ decays:
\beq
{\rm Re}\left( \frac{\epsilon ^{\prime}}{\epsilon ^{\prime}}
\right) =
\left\{
\begin{array}{l}
(2.80 \pm 0.41)\cdot 10^{-3} \; \; \; {\rm KTeV}
\cite{KTEV} \\
(1.40 \pm 0.43)\cdot 10^{-3} \; \; \; {\rm NA48}
\cite{NA48}
\end{array}
\right.
\eeq
however its exact size is still uncertain.
It is a discovery of the first rank irrespective of what
theory does or does not say.

Our theoretical interpretation of the
data is very much in limbo. As I had argued before a rather small, but
nonzero value is a natural expectation of the KM ansatz.
To go beyond such a qualitative statement, one has to evaluate
hadronic matrix elements; apparently one had
underestimated the complexities in this task.
One intriguing aspect in this development
is the saga of the $\Delta I=1/2$ rule: formulated in a compact way
\cite{DELTARULE} it
was originally expected to find a simple dynamical explanation; several
enhancement factors were indeed found, but the observed enhancement
could not be reproduced in a convincing manner; this problem was then
bracketed for some future reconsideration and it was argued that
$\epsilon ^{\prime}/\epsilon$ could be predicted while ignoring the
$\Delta I=1/2$ rule. Some heretics -- `early' ones
\cite{SANDA} and `just-in-time' ones \cite{TRIESTE} --
however argued that only approaches that reproduce the observed
$\Delta I=1/2$ enhancement can be trusted to yield a re alistic
estimate of $\epsilon ^{\prime}/\epsilon$. In particular
it had been suggested that the $\sigma$ -- the scalar $\pi\pi$ resonance
we have been hearing a lot of at this meeting --
plays a significant role here \cite{SANDA}.

In all fairness one
should point out that due to the large number of contributions with
different signs theorists are facing an unusually complex
situation. One can hope for
lattice QCD  to come through, yet it has to go beyond the
quenched approximation, which will require more time.

Direct CP violation will emerge also in hyperon decays. One can compare
partial widths for CP conjugate channels. Yet their differences are
given by the interference between $\Delta I=1/2$ and
$\Delta I=  3/2$ amplitudes
and thus suppressed by the $\Delta I=1/2$ enhancement similar to the
situation with $\epsilon ^{\prime}$ in $K_L \to \pi \pi$.
The HyperCP experiment instead compares the angular distributions
of protons and antiprotons in the decay sequences
$\Xi ^-\to \Lambda \pi ^- \to p \pi ^-\pi ^-$ and
$\bar \Xi ^+\to \bar \Lambda \pi ^+ \to \bar p \pi ^+\pi ^+$,
respectively \cite{RAMBO}.
An asymmetry there depends on the interference
between S- and P-waves; it is therefore not reduced by the
$\Delta I=1/2$ rule and could conceivably exceed the $10^{-4}$
level. HyperCP aims for a statistical accuracy of $1.4\cdot 10^{-4}$
in the combined asymmetry parameter ${\cal A}_{\Xi \Lambda}$; so far they
have achieved:
\beq
{\cal A}_{\Xi \Lambda} = (-1.6 \pm 1.3 \pm 1.6)\cdot 10^{-3}
\eeq

The KAMI experiment proposed for FNAL\cite{KAMI} aims at measuring the
branching ratio for $K_L \to \pi ^0 \nu \bar \nu$ which can proceed
only due to CP violation. Within the SM with the CKM implementation
of CP violation one predicts
\beq
{\rm BR}(K_L \to \pi ^0 \nu \bar \nu) = (3.1 \pm 1.3) \cdot 10^{-11}
\eeq
Such a project has to overcome daunting experimental challenges.
On the other hand it presents a highly intriguing theoretical perspective:
it would provide an independant route for extracting
sin$2\phi [\beta]$. Similar plans are being examined at BNL.

KTeV and NA48 will analyze various rare $K_L$ and $K_S$ modes
\cite{BELL,MUNDAY}.
Some are directly relevant to the CP phenomenology while others
build a broader data base for interpreting future measurement.
Using the literary figure of the "Cathedral Builders' paradigm"
I might add the following comment about the latter class of measurements:
while only engineers might perceive their beauty,
they are important elements for the foundations and thus for the
whole building.

\subsection{$\Delta B \neq 0$}

The second new element in 1999 was the start-up of the
new asymmetric $B$ factories BaBar and BELLE.
Their first results leave us
in limbo:
\bea
{\rm sin}2\phi _1[\beta] &=& 0.45 ^{+0.43 + 0.07}_{-0.44 - 0.09}  \; \;
{\rm BELLE} \\
{\rm sin}2\phi _1[\beta] &=& 0.12 \pm 0.37 \pm 0.09  \; \;
{\rm BaBar}
\eea
to be compared with the earlier data \cite{CDF}
\beq
{\rm sin}2\phi _1[\beta] = 0.79 ^{+0.41}_{-0.44} \;
{\rm CDF}
\eeq
No definite conclusions can be drawn yet. However the amazingly speedy
start-up of the two asymmetric beauty factories and the new runs of
D0 and CDF should should produce quite specific results within the
next two years \cite{D0,CDF}.

Nevertheless it is tempting to speculate `what if no asymmetry is
observed in $B\to \psi K_S$; i.e., if one finds, say,
$|$sin$2\phi [\beta]| \leq 0.05$. We would know there had to be New
Physics present since otherwise we could no longer accommodate
$K_L \to \pi \pi$. One would have to raise the basic question why the
CKM phase is so suppressed, unless there is an almost
complete cancellation
between KM and New Physics forces in $B\to \psi K_S$; this would shift
then the CP asymmetry in $B\to \pi \pi, \, \pi \rho$.

\subsection{Probing for New Physics}

As already stated, I expect that by 2002
CP violation will be stablished in at
least one $B$ decay mode, presumably in $B_d \to \psi K_S$.
This will be a discovery of the very first rank -- in all likelihood
the first observation of CP violation outside the $K_L$ system --
yet that will be far from the end of it!

Experiments at the upgraded $B$ factories at KEK and
SLAC together with new experiments at the LHC -- LHC-B -- and at
FNAL -- BTeV -- are expected to achieve experimental accuracies of
a few percent, and they will measure many more observables.
At the same time I expect that over the next five years or so
we will be able to predict Standard Model effects with a few percent
accuracy due to improved theoretical tools and new
measurements of CP {\em in}sensitive rates.
This will provide us with high sensitivity probes for the
presence of New Physics.

There are precedents for establishing the presence of
New Physics in such an {\em indirect} way in heavy flavour
decays: based on the apparent absence of flavour changing neutral
currents some courageous souls
\cite{GIM} postulated the existence of charm quarks;
the occurance of $K_L \to \pi \pi$ lead to the conjecture that
even a third family of quarks had to exist \cite{KM}. However in
all those cases we could rely on a {\em qualitative}
discrepancy; i.e., the difference between observed and predicted rate
amounted to several orders of magnitude or the predicted
rate was zero -- as for $K_L \to \pi \pi$.

 In the decays of beauty
hadrons we predict many large or at least sizeable effects, and
realistically in most cases we can expect differences well below an
order of magnitude only! Thus we will
have to deal with a novel challenge not encountered before; it will
require that we gain quantitative control over that most evasive
class of entities -- theoretical uncertainties. I am confident
we will make great progress in that respect. My optimism is
not based on hoping that novel
theoretical breakthroughs will occur -- they might.
But what will empower us is the
fact that so many different types of observables can be measured
in beauty decays. There are actually six KM unitarity triangles
\cite{TRIANGLES}, and several of their angles can be
measured in the dedicated and comprehensive research program
that is being undertaken world-wide. Our analysis will then
be able to invoke overconstraints -- the most effective weapon
in our arsenal against systematic uncertainties in general.

\section{The Rio Manifesto -- "I have come to praise c(harm),
not to bury it"}

A manifesto usually consists of three parts, namely (i) an analysis
of past history, (ii) lessons derived for the present and
(iii) a plan of action for the future. Furthermore it is usually
expressed with great conviction or at least great passion.
In my evaluation of charm physics I will largely, but not completely
follow this pattern: I will describe the past contributios of charm
physics and point out its potential for leading to fundamentally new
insights in the future. However, I will refrain from even attempting
to sketch a concrete plan for future action. Yet I have to admit that
I feel a certain passion about these issues. I have made this part as
self-contained as possible.

\subsection{Charm in the present Pantheon of
fundamental physics \cite{ROSNER}}

The existence of charm quarks had been postulated in 1964 (the year CP
violation was discovered) mainly to complete the quark lepton
correspondance, namely to match the four lepton flavours --
electrons, muons and their respective neutrinos -- with four
quark flavours \cite{BJ}. In 1970 a more specific motivation was added,
namely to suppress flavour changing neutral currents
\cite{GIM}. I should add, though, that these arguments were not
embraced by everybody; a not untypical attitude at that time
is expressed in the quote: "Nature is smarter than Shelley and can do
without charm".

In 1971 charm hadrons were first seen in cosmic ray data by Niu and his
group at Nagoya University
\cite{NIU}, then in 1974 by the HPW collaboration
through dimuons in deep inelastic neutrino scattering; hidden charm
emerged in the $J\psi$ resonance that gave rise to the
`October revolution' of 1974, before open charm was identified at
SPEAR in 1976.

After that things were never the same again:
\begin{itemize}
\item
It became the standard paradigm to describe subnuclear events in
terms of quark-gluon degrees of freedom where the latter were now
seen as quite physical objects rather than merely mathematical
entities.
\item
Postulating new quark flavours turned into a
popular sport, at least for some time.
\item
Theorists grew more confident in tackling the description of
nonleptonic weak decays, at least for charm hadrons.

\end{itemize}
One very early fruit of this change in attitude was the KM paper
\cite{KM} on CP violation. It postulated, as one of its scenarios
for implementing CP violation, at least three families when
only 1.5 families had been widely accepted in the community.
This jump was helped
considerably by two convictions held at Nagoya University at that time,
but not at many other places: due to
Sakata and his school, the notion of quarks as
real rather than merely mathematical had been
readily accepted, as had been the existence of charm due to the
discovery of Niu \cite{NIU}.

Soon it was conjectured that due to the charm quark mass exceeding
ordinary hadronic scales, strong interaction effects should become
more tractable in total lifetimes as well as
nonleptonic two-body decay modes
and in lepton spectra of semileptonic decays of charm hadrons.

With charm being the second member of a quark family its
Standard Model weak phenomenology becomes quite predictable and a
bit on the dull side as explained in more detail later. It is
typically viewed as the `St. John the Baptist' of heavy
flavour physics, i.e., the precursor of greater things to come. It
indeed filled this role admirably. There were three {\em experimental}
achievements in that vein:
\begin{itemize}
\item
Microvertex technology was developed to resolve secondary
decay vertices corresponding to lifetimes of order $10^{-13}$ sec.
This technology has turned into a central tool for studying
beauty decays with lifetimes of order $10^{-12}$ sec
(and lower $\gamma$ values).
\item
Various flavour tagging techniques have been developed, namely
both `same-side' flavour tagging -- $D^* \to D\pi$ -- as well
as `opposite-side' flavour tagging --
$\bar H_c D \to (l^-/K^+X)_{\bar H_c}D$; the corresponding
techniques have become essential in studying $B-\bar B$ oscillations
and CP violation.
\item
The concept of turning $e^+e^-$ `sweatshops' into `factories' has been
pioneered for charm with $e^+e^- \to J/\psi$ and
$e^+e^- \to \psi ^{\prime \prime} \to D \bar D$ pointing the
way to $e^+e^- \to \Upsilon (1S-3S)$ and
$e^+e^- \to \Upsilon (4S) \to B \bar B$.

\end{itemize}
New insights were gained also on the {\em theory} side:
\begin{itemize}
\item
Basic phenomenological concepts like
the so-called `Pauli Interference' (PI) \cite{RUCKL} or
`Weak Annihilation' (WA) \cite{BBS} were developed for treating
weak lifetimes of charm {\em hadrons}.
\item
Likewise for exclusive nonleptonic two-body channels: the
concepts of factorization, class I, II and III transitions
\cite{STECH}.
\item
Certain aspects of heavy quark symmetry were glimpsed first
in the charm system.
\item
The role of $SU(3)_{Fl}$ symmetry and its limitations in
exclusive versus inclusive transitions were studied.
\item
The $AC^2M^2$ model for semileptonic decay {\em spectra}
was first put forward \cite{ACCMM}.
\item
It was realized that charm physics provides a wide and
novel area for lattice QCD.

\end{itemize}
These considerations later were applied to the dynamics
of beauty hadrons with $1/m_Q$ expansions providing justification.

\subsection{Present profile of charm dynamics}

There is a triple motivation for heavy flavour studies in general,
and it applies to charm in particular:
\begin{itemize}
\item
It provides a novel probe of the {\em strong} interactions in charm
production, hadronic spectroscopy and the weak decays of charm
hadrons.
\item
One determines fundamental parameters, namely CKM parameters
and the charm quark mass.
\item
One searches for manifestations of New Physics in rare decays,
oscillations and CP asymmetries.

\end{itemize}

\subsubsection{QCD aspects}

{\bf (i) Charm production}

A large body of high quality data has been accumulated on photo- and
hadro-production of charm as reviewed by Appel \cite{APPEL}.
Distributions over a wide kinematical range have been studied,
including production asymmetries. The increase in statistics has lead
to a new level of sophistication:
\begin{itemize}
\item
Charm-anticharm correlations are analyzed in large samples where both
charm hadrons have been reconstructed.
\item
The polarizations of charm baryons and of charmonia are measured.

\end{itemize}
Alas, no {\em comprehensive} theoretical analysis has been
performed. I am not suggesting that a true treasure is
hidden there that will bring instant fame to the theorist
who uncovers it; however I do believe that this field can be
mined for valuable information.

Charm acts as a probe in two ways here: its mass provides a
(relatively) hard scale for the dynamics and thus a new handle on
the strong interactions and the hadronization driven by them.
At the same time charm production depends on the structure of the
beam and the target, namely their gluon and $q \bar q$ sea
distributions. Charm production thus can tag these contributions;
i.e., measurements of charm production in different reactions allows us
to extract the gluon and $q \bar q$ distributions.

Such studies provide
us with conceptual insights into the inner workings of QCD; it should
lead to a better understanding of the physical interpretation
underlying some fit parameters like `intrinsic' $k_T$ or the
notion of `colour drag'. Secondly
a good quantitative knowledge of gluon and $q \bar q$ distributions
represents `engineering' input into many other physics
studies in hadronic collisions. Finally a better understanding of
charm production is needed before it can safely be used as a probe
signalling the onset of the quark-gluon phase transition.

{\bf (ii) Spectroscopy}

There is a rich spectroscopy of the excitations of charm hadrons as
reviewed by Kutschke \cite{KUTSCHKE}. More surprising at first sight
are the contributions that the weak decay studies of charm hadrons
are making to the spectroscopy of hadrons carrying light flavour only.
Very detailed analyses of the Dalitz plots for
$D^+, D_s^+ \to \pi ^+\pi ^- \pi ^+$ and $D^+ \to K^- \pi ^+\pi ^+$
have been presented \cite{GOEBEL}. The emergence of the
$\sigma$ resonance in the $\pi \pi$ channel has been clearly
established. Of course, this resonance has been observed before
in low energy reactions with much larger statistics. The relevance here
does not lie in the statistical significance. The most remarkable point
is that the $\sigma$ emerges in a completely different environment with
consistent values for mass and width. I have already mentioned a more
surprising role played by the $\sigma$: it has been suggested as a
relevant factor driving the $\Delta I=1/2$ rule and
$\epsilon ^{\prime}/\epsilon$.

The fact that the $D_s$ lifetime exceeds that of $D^0$ by very close to
20 \%, see below, tells us that the {\em total}
$D_s$ widths receives a 10 - 20 \%
{\em destructive} contribution from WA. This leads to the obvious question:
Can one see the footprint of the latter in {\em exclusive}
channels? I find it intriguing that the $\sigma$ resonance, which manifests
itself clearly in $D^+ \to \pi ^+\pi^-\pi^+$, is absent in
$D_s^+ \to \pi ^+\pi^-\pi^+$, while both transitions are affected by
WA. These data presumably contain a quite relevant, albeit subtle
message which we have not decoded yet.

Lessons about other resonances are being learnt as well in Dalitz
plots like $D \to K \bar K \pi$ etc.: does the $K \bar K$ pair form
a resonance, a molecule or simply exhibit a threshold enhancement?

A more speculative idea is to look for glueball candidates $G$ in the
hadronic final state of semileptonic $D_s$ decays or in nonleptonic
$D_s$ (or Cabibbo suppressed $D^+$) decays once a fast pion has been
removed:
\bea
D_s [D^+] &\to& l^+ \nu G \\
D_s [D^+]&\to& \pi ^+_{fast} G
\eea

{\bf (iii) Lifetimes}

On general grounds one expects the following hierarchy
in lifetimes \cite{BELLINI,BILIC}:
\be
\tau (D^+) > \tau (D^0) \sim \tau (D_s^+) \geq
\tau (\Xi _c^+) > \tau (\Lambda _c^+) >
\tau (\Xi _c^0) > \tau (\Omega _c)
\label{PATTERNCHARM}
\ee
The $1/m_Q$ expansion can be applied to obtain some more
specific predictions which are listed in
Table \ref{TABLECHARM} and compared to the data.
\begin{table}
\begin{tabular} {|l|l|l|l|}
\hline
 & $1/m_c$ expect. \cite{BELLINI}& theory comments & data \\
\hline
\hline
$\frac{\tau (D^+)}{\tau (D^0)}$ & $\sim 2$ & PI dominant & $2.54 \pm 0.03 $
\cite{GOLUTVIN} \\
\hline
$\frac{\tau (D_s^+)}{\tau (D^0)}$ & 1.0 - 1.07 & {\em without} WA
\cite{DS}&
 $1.125 \pm 0.042$  PDG '98\\
 & 0.9 - 1.3 & {\em with} WA \cite{DS} & $1.18 \pm 0.02$ \cite{GOLUTVIN}
E791, CLEO, \\
 & $1.08 \pm 0.04$ & QCD sum rules
\cite{HY}&  FOCUS \\
\hline
$\frac{\tau (\Lambda _c^+)}{\tau (D^0)}$ & $\sim 0.5$
& quark model matrix  & $0.50 \pm 0.03$ PDG'00\\
 & & elements & \\
\hline
$\frac{\tau (\Xi _c^+)}{\tau (\Lambda _c^+)}$ & $\sim 1.3$ &
ditto & $1.60 ^{+0.30}_{-0.21}$ PDG'00\\
 & & & 2.14  FOCUS, very prelim. \\
\hline
$\frac{\tau (\Xi _c^+)}{\tau (\Xi _c^0)}$ & $\sim 2.8$ &
ditto & $3.37 ^{+0.98}_{-0.65}$ PDG'00 \\
 & & & 3.56 FOCUS, very prelim. \\
\hline
$\frac{\tau (\Xi _c^+)}{\tau (\Omega _c)}$ & $\sim 4$ &
ditto & $5.2 \pm 1.8$ PDG'00\\
\hline
\end{tabular}
\centering
\caption{Lifetime ratios in the charm sector}
\label{TABLECHARM}
\end{table}
A few comments are in order here:
\begin{itemize}
\item
There is agreement between expectations and the data even beyond
the qualitative level. This is
quite remarkable and not something that one could count on
for two reasons. (i) The longest and shortest lifetimes
differ by a factor of about twenty.
(ii) At the same time one should keep
in mind that the charm quark mass
exceeds ordinary hadronic scales by a moderate amount only; a
$1/m_c$ expansion will therefore
be sensitive to higher order terms that
cannot be calculated.
\item
In quoting a predicted lifetime ratio of about two
I am well aware that the measured value is different from two. Yet that
numerical difference is within the theoretical noise level: one could
use $f_D=220$ MeV rather than 200 MeV and WA, which has been
ignored here, could account for
10 - 20 \% of the $D^0$ width.
\item
PI is the main engine driving the $D^+ - D^0$ lifetime difference
as already anticipated in the `old' analysis of
Guberina et al. \cite{RUCKL}; the main impact of the HQE
for this point was to show that WA cannot constitute the leading
effect and that $BR_{SL}(D^0) \simeq 7 \%$,
$BR_{SL}(D^+) \simeq 17 \% $
 is consistent with PI being the
leading effect \cite{BELLINI}.

\item
Since $\tau (D_s)/\tau (D^0) \simeq 1.07$ can be generated {\em without} WA
\cite{DS},
the `old' data on $\tau (D_s)/\tau (D^0)$ had provided an independant
test for WA {\em not} being the leading source for
$\tau (D^0) \neq \tau (D^+)$; it actually allowed for it being quite
irrelevant. The `new' data reconfirm the first conclusion; at the same time
they point to WA as a still significant
process. This provides new impetus to uncover the impact of WA on
{\em exclusive} channels like semileptonic modes, 3-pion final states etc.,
as mentioned above.
The more ambitious analysis of Ref.\cite{HY} has yielded a value that
falls short of the recent measurements.

\item
The description of the {\em baryonic} lifetimes is helped by
forgiving experimental errors; more accurate measurements of
$\tau (\Xi _c^{+,0}, \Omega _c)$ might well
exhibit deficiencies in the theoretical description. We heard about
two new numbers from FOCUS, which are however still very preliminary
without a systematic error given.
\item
{\em Non}universal semileptonic widths --
$
\Gamma _{SL}(D) \neq \Gamma _{SL}(\Lambda _c) \neq
\Gamma _{SL}(\Xi _c) \neq \Gamma _{SL}(\Omega _c)
$ --
are predicted with the main effect being {\em constructive} PI in
$\Xi _c$ and $\Omega _c$ decays; the lifetime ratios among the
baryons will thus not get reflected in their semileptonic
branching ratios; one estimates \cite{VOLSL}
\bea
BR_{SL}(\Xi _c^0) \sim BR_{SL}(\Lambda _c)
\;  &\leftrightarrow& \;
\tau (\Xi _c^0) \sim 0.5 \cdot \tau (\Lambda _c) \\
BR_{SL}(\Xi _c^+) \sim 2.5 \cdot BR_{SL}(\Lambda _c)
\;  &\leftrightarrow& \;
\tau (\Xi _c^+) \sim 1.3 \cdot \tau (\Lambda _c) \\
BR_{SL}(\Omega _c)  \; &<& 15 \;  \%
\eea
Unfortunately nothing is known about them experimentally.
\end{itemize}

{\bf (iv) Nonleptonic two-body modes}

Very early on it had been conjectured that even the nonleptonic
two-body decays of charm hadrons could be described. Including
perturbative QCD renormalization of the $\Delta C=1$ operator
leads to the emergence of two multiplicatively renormalized
operator $O_{\pm}$ with short-distance coefficients $c_{\pm}$.
Using this
effective weak Lagrangian
and some simple prescription for evaluating hadronic matrix elements
it was predicted that
BR$(D ^0 \to K^0 \pi ^0) \ll {\rm BR}(D ^0 \to K^- \pi ^+)$
and BR$(D ^0 \to K^+ K ^-) \sim {\rm BR}(D ^0 \to \pi ^+ \pi ^-)$
should hold. Early data showed already that these predictions were
quite wrong
\footnote{This was the first sign that the decays of
charm hadrons were not that simple. Subsequently the first evidence
appeared that contrary to expectation
the $D^+$ and $D^0$ lifetimes differed by a very sizeable
factor; that factor appeared to exceed five initially before it
settled down to below three.}: the $SU(3)$ breaking in
$D^0 \to K^+ K^-$ versus $D ^0 \to \pi ^+ \pi ^-$ was found to be
very large and the expected suppression of $D ^0 \to K^0 \pi ^0$
did not materialize. People then `rediscovered' that charm decays
proceed in an environment populated by many resonances which will
affect exclusive modes significantly.

Stech and collaborators made the bold ansatz to describe
the nonleptonic two-body charm decays in terms of two
effective quantities $a_{1,2}$ which contained both the short-distance
coefficients $c_{\pm}$ and a parameter $\xi$ reflecting the evaluation
of the relevant hadronic matrix elements:
$a_1 = (c_+ + c_-)/2 + \xi (c_+ - c_-)/2$,
$a_2 = (c_+ - c_-)/2 + \xi (c_+ + c_-)/2$. In the valence quark
approximation and treating colour degrees of freedom statistically one
finds $\xi = 1/N_C = 1/3$.

They also classified the
channels into three categories:
\bea
{\rm class \; I}&:& \; \; D^0 \to M^+ M^{\prime -} \\
{\rm class \; II}&:& \; \; D^0 \to M^0 M^{\prime 0} \\
{\rm class \; III}&:& \; \; D^+ \to M^+ M^{\prime 0}
\eea
with class I[II] transitions being described by $a_1$ [$a_2$] only;
class III channels depend on the interference between
$a_1$ and $a_2$ terms.

It is quite amazing that a very decent fit could be obtained.
For while the ansatz allows for final state interactions, it does
so only in a `universal' sense by lumping them together into
the two fit parameters $a_{1,2}$.  Even
more intriguingly the values found
for $a_{1,2}$ were quite consistent with what one obtains when
only terms leading in $1/N_C$
(with $N_C$ denoting the number of colours) are retained in the
prescription for the matrix elements
\cite{BURASNC}. Yet
ultimately, efforts to gain
some deeper understanding through $1/N_C$ expansions did not
succeed.

There are several good reasons why the QCD treatment sketched above
for nonleptonic beauty decays cannot be applied here; for example,
contributions of order $1/m_Q$ that cannot be treated in that scheme
are presumably very large here. Nevertheless one should have a look
at it.

In any case, these so-called BSW-type
studies were and still are very relevant exercises:
\begin{itemize}
\item
They provide us with some new insights into hadronisation at the
interface of perturbative and nonperturbative dynamics.
Since these two-body modes make up the bulk of all nonleptonic
charm decays, they can teach us novel lessons on quark-hadron duality.
\item
Such lessons are also quite unique. For
nonleptonic two-body modes in $B$ decays being much less frequent
and prominent can teach us little about
duality by themselves. Also class II transitions, often referred to
as colour suppressed, are hard to study there because of their
tiny branching ratios.
\item
These studies constitute an essential piece in theoretical
engineering in understanding $D^0 - \bar D^0$ oscillations and
CP violation once they have been observed. I will return to this
point later on.

\end{itemize}

{\bf (v) $D \to l \nu$}

$D_s$ decays into a purely leptonic final state have been found
by several experiments yielding an extraction of the
decay constant, as stated above in Eq.(\ref{FDS}). At this
point I want to stress the particular role played by charm:
(i) The branching
ratio is much larger for charm than for beauty mesons and thus
more accessible experimentally.
(ii) At the same time the decay constant is also more
accessible to lattice QCD for the charm than the beauty
mass scale. To have control over the theoretical uncertainties,
one has to go to a fully unquenched treatment, i.e. where
all light flavours are treated {\em dynamically}. Charm studies
have now yielded the first partially unquenched results for
$f(D)$, namely with two light dynamical flavours.
(iii) The two items listed above concern the decay constants as
a measure of our theoretical control over QCD. On top of that
the decay constants for $B$ and $B_s$ mesons represent also
an essential engineering input for interpreting the
strength of $B^0 - \bar B^0$ oscillations.
Of course, one wants to measure $f(B)$ directly.
Yet that will be quite a challenge experimentally on top of the
fact that $B \to \mu \nu$ depends on $|V(ub)|$ as well.
On the other hand if the measured values of
$f(D)$ and $f(D_s)$ agree with complete lattice results,
this would give us great confidence in the extrapolation of
lattice results to the beauty scale.

{\bf (vi) $D \to l \nu P/V$}

One studies $D \to l \nu K/\pi$ and $D \to l \nu K^*/\rho$
transitions as a check for the theoretical control we have
achieved over hadronization effects. Again the motivation is
two-fold, namely to check our mastery over QCD and to apply this
knowledge to semileptonic $B$ decays where one wants to extract
$|V(cb)|$ and $|V(ub)|$. The charm system has the advantage
that it is more accessible to {\em un}quenched lattice studies.

\subsubsection{Weak dynamics}

{\bf (i) Tree level transitions}

The full Cabibbo hierachy has been observed now, namely
Cabibbo allowed and suppressed channels in semileptonic
transitions and Cabibbo allowed, suppressed and doubly
suppressed ones in nonleptonic decays \cite{BOCA}.

Imposing three-family unitarity one has, see Eq.(\ref{CKM3}):
\beq
|V(cs)| = 0.9742 \pm 0.0008 \; , \; |V(cd)| = 0.222 \pm 0.003
\eeq
Yet without that constraint the values are considerably less
precise, see Eq.(\ref{CKM4}):
\beq
|V(cs)| = 0.880 \pm 0.096 \; , \; |V(cd)| = 0.226 \pm 0.007
\eeq
As far as $|V(cs)|$ is concerned the main information
from semileptonic $D$ decays is augmented by findings from charm
production in deep inelastic neutrino scattering; for
$|V(cd)|$ it is the other way around. A recent
OPAL analysis \cite{OPAL}  of $W\to$ charm jets obtains
\beq
|V(cs)| = 0.969 \pm 0.058
\eeq

With charm production and decay rates depending on the
charm quark mass, the latter can be extracted from a measurement
of those. However this represents a much more involved task
than a simple parton model description would suggest. For in a
quantum field theory the value of a mass depends not only on the
scale at which it is evaluated, but also on how it is defined
(and renormalized): does one use a pole or a $\overline{MS}$ or a
`kinetic' mass? Since quarks are confined, there is no natural
definition like for electrons. While the pole mass for charm quarks
has many convenient features, it suffers from an irreducible theoretical
uncertainty of order $\Lambda _{QCD}$. The $\overline{MS}$ mass
is quite appropriate for production processes well above threshold where
typical momenta are large compared to $m_c$, yet ill-suited
for charm decays where momenta are below $m_c$. In that domain
a mass like the `kinetic' one is much better-suited since it
has a softer infrared behaviour \cite{HQT}. $\overline{MS}$ and kinetic mass
are related, but do not coincide even at the same scale.

{\bf (ii) Loop transitions}

It is often stated that rare charm decays are extremely rare,
$D^0$ oscillations are slow and
CP asymmetries tiny within the SM and that therefore their analysis
provides us with zero-background searches for New Physics.

With the enhanced experimental sensitivity that has been
or will be achieved the real question is:
"How slow is slow and how tiny is tiny?"

Flavour changing neutral current transitions
$D\to l^+l^-$, $D\to \gamma X$ and $D \to l^+l^-X$
can be generated by effective
local operators due to loop effects in analogy to rare $B$ decays.
However the resulting branching ratios are absolutely
minuscule, even after some huge enhancement due to radiative
QCD corrections. Furthermore -- and even worse -- long distance
dynamics induces effects that are larger by orders of magnitude
\cite{HEWETT}.

Oscillations are described by the normalized mass and width
differences:
$x_D \equiv \frac{\Delta M_D}{\Gamma _D}$,
$y_D \equiv \frac{\Delta \Gamma}{2\Gamma _D}$.
A very {\em conservative} SM estimate yields
\beq
x_D \, , \; y_D \leq {\cal O}(0.01)
\label{CONSERV}
\eeq
reflecting mainly the fact that the
amplitudes for channels communicating between $D^0$ and $\bar D^0$
are proportional to tg$\theta _C^2 \simeq 0.05$.

The flavour tag in the final state is provided by the charge of the
lepton or the kaon in $D^0 \to l^+ \nu K^-$ or
$D^0 \to K^- \pi 's$, respectively. Within the SM with its
$\Delta Q = - \Delta C$ rule semileptonic decays provide a
clean handle:
\beq
\frac{\Gamma (D^0 \to l^-X)}{\Gamma (D^0 \to l^+X)}\equiv
r_D \simeq \frac{x_D^2 + y_D^2}{2}
\eeq
which can be searched for in time integrated as well as time resolved
rates:
\beq
\frac{d\Gamma (D^0 \to l^-X)}{dt} \propto e^{-\Gamma t} \cdot  r_D
\cdot (\Gamma t)^2
\eeq
When using kaons in nonleptonic decays one has to contend with the
background from doubly Cabibbo suppressed channels (DCSD)
leading to a more complex time evolution
\beq
\frac{d\Gamma (D^0 \to K^+\pi 's)}{dt} \propto e^{-\Gamma t}
\left[ r_D (\Gamma t)^2 + |\hat T_{DCSD}|^2 +
|\hat T_{DCSD}| y^{\prime} \Gamma t\right]
\eeq
where
\beq
\hat T_{DCSD} = \frac{T(D^0 \to K^+ \pi ^-)}{T(D^0 \to K^- \pi ^+)}
\eeq

The experimental landscape is described by the following numbers:
\bea
x_D &\leq& 0.03   \\
 y_D &=&
\left\{
\begin{array}{l}
(0.8 \pm 2.9 \pm 1.0) \% \; \; {\rm E791} \\
(3.42 \pm 1.39 \pm 0.74) \% \; \; {\rm FOCUS} \\
(1.0^{+3.8 + 1.1}_{-3.5-2.1})\% \; \; {\rm BELLE}
\end{array}
\right.
\\
y_D^{\prime} &=& (-2.5 ^{+1.4}_{-1.6} \pm 0.3)\% \; \; {\rm CLEO}
\eea
E 791 and FOCUS compare the lifetimes for two different channels,
whereas CLEO fits a general lifetime evolution to
$D^0(t) \to K^+\pi ^-$; its $y_D^{\prime}
= -x_D{\rm sin}\delta + y_D {\rm cos}\delta $ depends on the strong
rescattering phase $\delta$ between $D^0 \to K^-\pi^+$ and
$D^0 \to K^+\pi^-$ and therefore could differ substantially from
$y_D$ if that phase were
sufficiently large \cite{PETROV}.
All measurements are still consistent with zero.

The SM with the KM ansatz allows {\em direct}
CP asymmetries in time integrated
partial widths to emerge for singly Cabibbo suppressed modes. A
benchmark guestimate can be inferred from considering just CKM
parameters \cite{TAUCHARM}:
\beq
{\rm asymmetry} \sim {\cal O}(\lambda ^4) \sim {\cal O}(10^{-3})
\eeq
The analysis of Ref.\cite{LUSIG} based on theoretical engineering
concerning hadronic matrix elements and strong phase shifts
finds somewhat smaller numbers.

Data are summarized in Table \ref{CPAS} \cite{PEDRINI}.
\begin{table}
\begin{tabular} {|l|l|l|l|}
\hline
 Decay mode& $D^0 \to K^+ K^-$ & $D^0 \to \pi^+ \pi^-$ &
$D^{\pm} \to K^+ K^-\pi ^{\pm}$ \\
\hline
\hline
E 791 & $- 1.0 \pm 4.9 \pm 1.2$\% & $- 4.9 \pm 7.8 \pm 3.0$\%
 & $- 1.4 \pm 2.9$\%
 \\
\hline
CLEO & $ 0.04 \pm 2.18 \pm 0.84$\% &
$ 1.94 \pm 3.22 \pm 0.84$\% &
 \\
\hline
FOCUS & $- 0.1 \pm 2.2 \pm 1.5 $\%  &
$ 4.8 \pm 3.9 \pm 2.5 $\%  &
$ 0.6 \pm 1.1 \pm 0.5 $\%
 \\
\hline
\end{tabular}
\centering
\caption{Data on direct CP asymmetries in $D$ decays}
\label{CPAS}
\end{table}
The experimental sensitivity has increased significantly
to put us within striking distance of the 1\% level. Yet
the numbers are still consistent with zero and
we are still above the level expected for CKM effects.

\subsubsection{ First resume}

At this point one might subscribe to the following view.
There is certainly unfinished business:
\begin{itemize}
\item
The accuracy with which {\em absolute} branching ratios are
known, in particular for $D_s$ and charm baryon decays, leaves
something to be desired;
\item
likewise for $\Xi _C^{0,+}$ and $\Omega _c$ lifetimes as well
as the semileptonic branching ratios of charm baryons.
\item
One wants to have measurements of $D\to l \nu$ and more precise
data on $D_s \to l \nu$.
\item
Post-MARKIII data on lepton {\em spectra} in
{\em inclusive} semileptonic charm decays would yield important
information on the hadronization process.
\item
One wants to enhance the experimental sensitivity for
$D^0 - \bar D^0$ oscillations and CP violation.
\end{itemize}
Such measurements would form important engineering inputs to
beauty studies and at the same time produce interesting lessons
on nonperturbative QCD dynamics. While the data indeed can be
improved and polished, one might be inclined to draw a somewhat cynical
summary: "Charm -- from a revolutionary to a petit bourgeois!"
Such evolutions do indeed often happen in history; I will however
argue that it is decidedly wrong here.

\subsection{Charm -- like Botticelli in the Sistine Chapel}

Instead I would characterize the situation
by referring to "charm as David Duval versus beauty as Tiger
Woods". Colleagues of mine who golf tell me that Duval was
widely considered the world's greatest golfer -- till
Woods appeared on the scene. In presenting my arguments
let me proceed from the general to the specific.
\begin{itemize}
\item
Charm quarks are the only up-type quarks that allow a full range of
indirect searches for New Physics:
\begin{itemize}
\item
As already mentioned $D^0 - \bar D^0$ oscillations can proceed even if
slowly; on the other hand there are no $T^0 - \bar T^0$ oscillations
since top quarks decay before they can hadronize \cite{DOK}.
\item
Likewise one can search for CP violation in $D^0 - \bar D^0$ oscillations,
whereas nothing like that can happen for top states.
\item
One can probe for direct CP violation in {\em exclusive} modes that
command decent branching ratios  in the charm case. In top decays
such branching ratios are truly tiny and the required coherence
is basically lost.
\item
Finally charm decays proceed in an environment populated with
many resonances which induce final state interactions (FSI) of
great vibrancy. While this feature complicates the interpretations
of a signal (or lack thereof) in terms of microscopic quantities, it
is optimal for getting an observable signal. In that sense it
should be viewed as a glass half full rather than half empty.

\end{itemize}
\item
Charm hadrons provide several practical advantages and opportunities:
\begin{itemize}
\item
Their production rates are relatively large.
\item
They possess long lifetimes.
\item
$D^* \to D\pi$ decays provide as good a flavour tag as one can have.

\end{itemize}

\end{itemize}
This leads to my basic contention: charm transitions are a
{\em unique} portal for obtaining a {\em novel} access to the
{\em flavour} problem with the
{\em experimental situation being a priori favourable}!

My discussion in no way aims at downgrading beauty physics -- my
intention is to point out the qualities of charm studies that
are easily overlooked in such a juxtaposition. This is already
indicated in the title of this subsection.

\subsubsection{$D^0 - \bar D^0$ oscillations}

Comparing the theoretical bound on $D^0 - \bar D^0$
oscillations given in
Eq.(\ref{CONSERV}) with the available data tells us that
{\em only now have experiments begun to probe a range of values
for $x_D$ and $y_D$ where there is a realistic chance for a
non-zero value!} For I find it quite unlikely that New Physics
could overcome the second order Cabibbo suppression.

More restrictive bounds than the one stated in Eq.(\ref{CONSERV})
have appeared in the literature. Usually it has been claimed
that the contributions from the
operator product expansion (OPE) are completely insignificant
and that long distance contributions {\em beyond} the OPE provide the
dominant effects yielding $x_D^{SM}$, $y_D^{SM}$
$\sim {\cal O}(10^{-4} - 10^{-3})$. A recent detailed analysis
\cite{BUOSC} revealed
that a proper OPE treatment reproduces also such long distance
contributions with
\beq
x_D^{SM}|_{OPE}, \, y_D^{SM}|_{OPE} \sim {\cal O}(10^{-3})
\eeq
and that $\Delta \Gamma $, which is generated from
on-shell contributions, is -- in contrast to $\Delta m_D$
-- basically
insensitive to New Physics while on the other hand more susceptible
to violations of (quark-hadron) duality.

The FOCUS data contain a suggestion that the lifetime
difference in the
$D^0 - \bar D^0$ complex might be as large as ${\cal O}(1\% )$.
{\em If} $y_D$ indeed were $\sim 0.01$, two scenarios could arise
for the mass difference. If $x_D \leq {\rm few} \times 10^{-3}$
were found, one would infer that the $1/m_c$ expansion yields a
correct semiquantitative result while blaming the large value for
$y_D$ on a sizeable and not totally surprising violation of
duality. If on the other hand $x_D \sim 0.01$ would emerge, we would face
a theoretical conundrum: an interpretation ascribing this to New
Physics would hardly be convincing since $x_D \sim y_D$. A more sober
interpretation would be to blame it on duality violation or on the
$1/m_c$ expansion being numerically unreliable. Observing
$D^0$ oscillations {\em per se} then would not constitute an unambiguous
signal for New Physics. Yet if indeed $y_D \geq 0.01$
were established, it would represent undoubtedly an important
discovery even if the intrepretation were ambiguous.

Since the main motivation is to uncover New Physics, one should not treat
the SM $\Delta Q = - \Delta C$ rule as sacrosanct, but entertain the
notion that $e^{\Gamma t} d\Gamma (D^0(t) \to l^-X)/dt$ might
contain a genuine time independant contribution.

\subsubsection{CP violation involving $D^0 - \bar D^0$ oscillations}

The interpretation is much clearer once one finds a CP
asymmetry that involves oscillations; i.e., one compares
the time evolution of transitions like $D^0(t) \to K_S \phi$,
$K^+ K^-$, $\pi ^+ \pi ^-$ and/or
$D^0(t) \to K^+ \pi ^-$ with their CP conjugate channels. A
difference for a final state $f$ would depend
on the product
\beq
{\rm sin}(\Delta m_D t) \cdot {\rm Im} \frac{q}{p}
[T(\bar D\to f)/T(D\to \bar f)] \; .
\eeq
I want to stress two
aspects of this expression:
\begin{itemize}
\item
With  both factors being
$\sim
{\cal O}(10^{-3})$ in the SM with the KM ansatz
one predicts a practically zero
asymmetry $\leq 10^{-5}$.
\item
New Physics could quite conceivably generate
considerably larger values, namely
$x_D \sim {\cal O}(0.01)$,
Im$\frac{q}{p}
[T(\bar D\to f)/T(D\to \bar f)] \sim  {\cal O}(0.1)$
leading to an asymmetry of ${\cal O}(10^{-3})$.
\item
One should note that the
oscillation dependant term is linear in the
small quantity $x_D$
\beq
{\rm sin}\Delta m_D t
\simeq x_D t /\tau _D
\eeq
in contrast to $r_D$ which is
quadratic:
\beq
r_D \equiv \frac{D^0 \to l^-X}{D^0 \to l^+X}
\simeq \frac{x_D^2 + y_D^2}{2}
\eeq
It would be very hard to see $r_D = 10^{-4}$ in CP insensitive
rate. It could then well happen that $D^0 - \bar D^0$
oscillations are first discovered in such CP asymmetries!

\end{itemize}

\subsubsection{Direct CP violation}

{\bf (i) Partial widths}

There are three requirements for an asymmetry to become observable
between CP conjugate partial widths, namely (i) two coherent
amplitudes with (ii) a relative {\em weak} phase and (iii) a nontrivial
strong phase shift.

In Cabibbo favoured as well as in doubly Cabibbo suppressed
channels those requirements can be met with New Physics only. There is
one exception to this general statement: the transition
$D^{\pm} \to K_S \pi ^{\pm}$ reflects the interference between
$D^{+} \to \bar K^0 \pi ^+$ and $D^+ \to K^0 \pi ^+$ which
are Cabibbo favoured and doubly Cabibbo suppressed, respectively.
Furthermore in all likelihood those two amplitudes will exhibit
different phase shifts since they differ in their isospin
content \cite{YAMA}.

There is one effect that has to be there without any
theory uncertainty and without New Physics, namely an
asymmetry driven by the CP impurity in the $K_S$ state:
\beq
\frac{\Gamma (D^+ \to K_S \pi ^+) - \Gamma (D^- \to K_S \pi ^-)}
{\Gamma (D^+ \to K_S \pi ^+) + \Gamma (D^- \to K_S \pi ^-)} =
-2{\rm Re}\epsilon _K \simeq - 3.3 \cdot 10^{-3}
\label{DKSSM}
\eeq
In that case the same asymmetry both in magnitude as well
as sign arises for the experimentally much more challenging
final state with a $K_L$.
If on the other hand New Physics is present in $\Delta C=1$ dynamics,
most likely in the doubly Cabibbo transition, then both the
sign and the
size of an asymmetry can be different from the number in Eq.(\ref{DKSSM}),
and by itself
it
would make a contribution of the opposite sign to the asymmetry in
$D^+ \to K_L\pi ^+$ vs. $D^- \to K_L\pi ^-$.

Searching for {\em direct} CP violation in
Cabibbo suppressed $D$ decays as a sign for New Physics would also
represent a very complex challenge: within the KM description one expects
to find some asymmetries of order 0.1 \%
\cite{LUSIG}; yet it would be hard
to conclusively rule out some more or less accidental enhancement due to a
resonance etc. raising an asymmetry to the 1\% level.
Observing a CP
asymmetry in charm decays would certainly be a first rate discovery even
irrespective of its theoretical interpretation, as it is with respect
to $\epsilon ^{\prime}/\epsilon$. Yet to make a case that a signal in a singly
Cabibbo suppressed mode reveals New Physics is quite iffy. In all
likelihood one has to analyze at least several channels with comparable
sensitivity to acquire a measure of confidence in one's interpretation.

{\bf (ii) Final state distributions}

For channels with two pseudoscalar mesons or a pseudoscalar and a vector
meson a CP asymmetry can manifest itself only in a difference between
the two partial widths, as just discussed. If, however, the final state
is more complex -- being made up by three pseudoscalar or two
vector mesons etc. -- then it contains more dynamical information than
expressed by its partial width and CP violation can emerge also through
asymmetries in final state distributions. One general comment
still applies: since also such CP asymmetries require the
interference of two weak amplitudes, within the SM
they can occur in Cabibbo
suppressed modes only.

In the simplest such scenario one compares CP conjugate
{\em Dalitz plots}. I have noticed that Dalitz plot studies are
very popular with our colleagues in Rio de Janeiro. Having come here
I understand why: the topography of Rio with its steeply rising
mountain ranges crowned by spectacular peaks and separated by narrow
valleys is a dramatic large scale model of a Dalitz plot.

In the present context I want to stress the following: it is quite
possible that different regions of a Dalitz plot exhibit CP
asymmetries of varying signs that largely cancel each other when
one integrates over the whole phase space. I.e., subdomains of the
Dalitz plot could contain considerably larger CP asymmetries
than the integrated partial width.

Once a Dalitz plot is fully understood with all its resonance and
non-resonance contributions including their strong phases, one has a
powerful and sensitive new probe. This is not an easy goal to
achieve, though, in particular when looking for effects that
presumably are not overly large. It might be more promising
as a practical matter to start out with a more euristic approach. Let me
remind you of a quote by someone who often appears to be one of
the founding fathers of the American school of philosophy,
namely Yogi Berra
\footnote{Interestingly enough he practised Einstein's dictum that
one should not do one's deep thinking in one's paid job.}; he once
declared: "You can always start an observation of something by looking
at it!" My point is that we have still to learn the tricks of the trade;
this is best done by analysing data to see what can happen.
While some relevant experience exists from analyses of
$K \to 3\pi$, the situation is considerably more involved in
charm decays where the much larger phase space allows for many
resonances to make their presence felt.

I therefore suggest to undertake one's search for an asymmetry in an
open-minded fashion. One simple strategy would be to focus on an area
with a resonance band and analyze the density in stripes {\em across} the
resonance as to whether there is a difference in CP conjugate plots.

For more complex final states containing
four pseudoscalar mesons etc. other probes have to be
employed.  Consider for example
\beq
D^0 \to K^+K^- \pi ^+ \pi ^- \; ,
\eeq
where one can form a T-odd correlation with the momenta:
\beq
C_T \equiv \langle \vec p_{K^+}\cdot
(\vec p_{\pi^+}\times \vec p_{\pi^-})\rangle
\eeq
Under time reversal T one has
\beq
C_T \to - C_T
\eeq
hence the name `T-odd'. Yet $C_T \neq 0$ does not necessarily
establish T violation. Since time reversal is implemented
by an {\em anti}unitary operator, $C_T \neq 0$ can be induced by
final state interactions (FSI). While in contrast to the situation
with partial width differences FSI are not required to produce
an effect, they can act as an `imposter' here, i.e. induce a T-odd
correlation with T-invariant dynamics. This ambiguity can unequivoally
be resolved by measuring
\beq
\bar C_T \equiv \langle \vec p_{K^-}\cdot
(\vec p_{\pi^-}\times \vec p_{\pi^+})\rangle
\eeq
in $\bar D^0 \to K^+K^- \pi ^+ \pi ^- $; finding
\beq
C_T \neq - \bar C_T
\eeq
establishes CP violation without further ado!

Decays of {\em polarized} charm baryons provide us with a
similar class of observables; e.g., in
$\Lambda _c \Uparrow \; \to p \pi ^+\pi ^-$, one can analyse the
T-odd correlation $\langle \vec \sigma _{\Lambda _c}
\cdot (\vec p_{\pi ^+} \times \vec p_{\pi ^-})\rangle$ \cite{BENSON}.

\subsubsection{Benchmark numbers}

Without a clearcut theory of New Physics one has to strike a balance
between the requirements of feasibility and the demands of making a
sufficiently large step beyond what is known when suggesting
benchmark numbers for
the experimental sensitivity to aim at. In that spirit I suggest the
following numbers:
\begin{itemize}
\item
Probe $D^0 - \bar D^0$ oscillations down to
$x_D$, $y_D$ $\sim {\cal O}(10^{-3})$ corresponding to
$r_D \sim {\cal O}(10^{-6} - 10^{-5})$.
\item
Search for {\em time dependant} CP asymmetries in
$D^0(t) \to K^+K^-$, $\pi ^+\pi ^-$, $K_S\phi$ down to the
$10^{-4}$ level and in the doubly Cabibbo suppressed mode
$D^0(t) \to K^+ \pi ^-$ to the $10^{-3}$ level.
\item
Look for asymmetries in the partial widths for
$D^{\pm} \to K_{S[L]}\pi ^{\pm}$ down to $10^{-3}$ and likewise
in a {\em host} of singly Cabibbo suppressede modes.
\item
Analyze Dalitz plots and T-odd correlations etc. with a sensitivity
down to ${\cal O}(10^{-3})$.

\end{itemize}

\subsection{The future: expectations, promises and dreams}

It is guaranteed that the data base on charm physics will be expanded
tremendously in the several next years: FOCUS and SELEX will
presumably present their full analysis not later than at the
next meeting in this series two years from now; the $e^+e^-$
beauty factories at KEK, SLAC and Cornell will produce a large
amount of high quality data over the next several years;
COMPASS will make contributions as well.

There is considerable promise that BTEV and LHC-b can harness
the large statistics
of charm decays occurring in hadronic collisions for further studies
that might achieve a new statistical quality.

Finally there are the `gleam in the eye' plans for novel initiatives:
a $\tau$-charm factory at Cornell, deep inelastic neutrino scattering
at $\nu$ factories and a glue-charm factory at GSI.

Studies of charm hadrons have had an illustrious past; they were
instrumental in the Standard Model gaining universal acceptance.
Yet they are not a closed chapter; they have the potential to point
the way to New Physics. This educated hope can be expressed in a phrase
borrowed from the political scene: {\em "Charm physics: You haven't
seen anything yet -- the best things are still to come!"}

\section{The outlook}

Advances do not occur in a continuous fashion. Long
periods of seemingly little progress are often followed
by short periods of rapidly happening new developments that
sometimes lead even to a new paradigm. As I have stated
in the beginning we have entered such a special period for heavy
quark flavour physics, in particular with respect to CP
violation. The existence of direct CP violation has been
established in $K_L$ decays, we are on the brink of observing
for the first time CP asymmetries in a different system, namely
in beauty decays, and the hope for finding such effects in charm
decays no longer has to be based on "irrational exuberance"
\footnote{To say that data had not yielded a qualitatively
new result for more than thirty years does not reflect a
deeper truth. New empirical insights typically are not born
like the goddess Athene who jumped fully developed and in full
armour out of Zeus', her father's, head. They require substantial
incubation periods.
}.
Experimental studies relying on refined theoretical tools might
well reveal an incompleteness of the Standard Model in the
foreseeable future. I would like to emphasize in this context that
no findings in beauty physics -- no matter whether they are positive
or negative, whether they reveal New Physics or not -- can eliminate
the need for a dedicated high sensitivity program on charm (and top)
transitions.
\footnote{
Rio is
experiencing a phenomenon of manifest symbolism for our agenda:
penguins are showing up at the beaches in record numbers. The
reasons underlying that migration are not clear. The penguins arrive in
poor shape needing care; apparently they come against their will, and
the people do not know what to do with them, see
http://www.sueddeutsche.de/news/pinguine.htm.}

One of the puzzles of the SM, namely why neutrinos are massless,
might be solved now: they are actually not massless since they
show signs of oscillations! Neutrinos being very light is naturally
understood through the `see-saw' mechanism. Such findings open the
gates to vast new domains of dynamics characterised by the
emergence of a leptonic analogue to the CKM matrix, namely a
nontrivial
MNS matrix for lepton flavours \cite{MNS}.  Effects will be
even more subtle than their quark analogues, and searches for them
will severely test our patience and dedication. Yet having a
family structure with its lepton-quark correspondances it makes
eminent sense for the flavour dynamics to stand on two legs, even
if one is stronger than the other. I am also extremely pleased that
the organizers of this series of meetings react substantially
to these developments by changing the name of the series to
`Heavy Quark and Lepton Flavours'. At the same time I would like to
urge the organizers to maintain the tradition of giving many
time slots to younger speakers (which means substantially
younger than yours truly).

Finally I would like to thank Alberto Reis and his team in Brazil
for creating such a wonderful atmosphere for the meeting and being
such gracious hosts. In addition I want to
express my deep appreciation and gratitude for their pioneering work at
CBPF and their
universities in creating a home for fundamental
reasearch in general and high energy physics in particular in the southern
hemisphere of our world!

\vskip 3mm
{\bf Acknowledgements}

This work has been supported by the NSF under the grant
PHY 96-05080.



\begin{thebibliography}{99}
\bibitem{CDF}
J. Boudreau, these Proceed.; it contains delightful illustrations
of $B^0 - \bar B^0$ oscillations and CP violation by optical
analogies.
\end{thebibliography}
\end{document}